\documentclass{emulateapj}

\begin{document}

\title{Spitzer Observations of Low Luminosity Isolated and Low Surface Brightness Galaxies}

\author{J. L. Hinz, M. J. Rieke, G. H. Rieke, C. N. A. Willmer, K. Misselt,
C. W. Engelbracht, M. Blaylock}
\affil{Steward Observatory, University of Arizona, 933 N. Cherry Ave.,  Tucson,
AZ  85721
\\email: jhinz, mrieke, grieke, cnaw, kmisselt, cengelbracht, blaylock@as.arizona.edu}

\author{T. E. Pickering}

\affil{MMT Observatory, Smithsonian Institution and University of Arizona, Tucson, AZ 85721
\\email:  tim@mmto.org}

\begin{abstract}

We examine the infrared properties of five low surface brightness galaxies
(LSBGs) and compare them with related but higher surface brightness
galaxies, using {\it Spitzer Space Telescope} images and spectra.
All the LSBGs are detected in the 3.6 and 4.5\,$\micron$ bands, 
representing the
stellar population.  All but one are detected at 5.8 and 8.0\,$\micron$, 
revealing emission from hot dust and aromatic molecules, though many are
faint or point-like at these wavelengths.  Detections of LSBGs 
at the far-infrared wavelengths, 24, 70, and 160\,$\micron$, are varied in 
morphology and brightness, with only two detections at 160\,$\micron$,
resulting in highly varied spectral energy distributions.
Consistent with previous expectations for these galaxies, we find that 
detectable dust 
components exist for only some LSBGs, with the strength of dust emission 
dependent on the existence of bright star forming regions.  However, the
far-infrared emission may be relatively weak compared with normal
star-forming galaxies.

\end{abstract}


\section{INTRODUCTION}
Low surface brightness galaxies (LSBGs), defined as those with $B$-band 
central surface brightnesses, $\mu_{0, B}$, fainter 
than 23.0 mag arcsec$^{-2}$, appear to have 
followed a very different evolutionary history from high surface brightness 
galaxies (e.g., McGaugh 1992; Knezek 1993).  Their
stellar populations, stellar masses, current star formation rates, and
other properties appear to differ significantly from their high surface
brightness counterparts, and little is known about their corresponding dust
properties.  A better understanding of these differences is needed to
understand how they evolved to their present state.

It has been generally assumed, even though a population of red LSBGs has 
been discovered (O'Neil et al. 1997a), that they contain little
dust, and dust reddening has not been considered an
important effect (e.g., Bothun et al. 1997; Bell et al. 2000).  
LSBGs have low star formation rates, with
suspected modest bursts in the
range 10$^{-3}$-10$^{-2}$\,M$_{\odot}$\,yr$^{-1}$ (e.g.,
Vallenari et al. 2005) and low metallicities, with almost all LSBGs at
or less than about one-third solar (McGaugh 1994; Ronnback \& Bergvall
1995).  Their low metallicities imply that the dust to gas ratios should
be systematically lower than in their high surface brightness galaxy
counterparts, and the fact that the 
{\it Infrared Astronomical Satellite} ({\it IRAS}; Neugebauer et al.\ 1984) 
and the {\it Infrared Space Observatory} ({\it ISO}; Kessler et al.\ 1996) were
only able to detect two of these galaxies seems to indicate that dust
is much less important in LSBGs.  Furthermore, any data acquired with these
observatories would not have been adequate to characterize dust temperatures or
spatial distributions.

That dust plays a relatively minor role in the evolution of LSBGs
is further reinforced by observations of highly transparent galaxies
that appear to have multiple distant galaxies seen through their
disks (O'Neil et al.\ 1997b; P.\ Knezek, private communication).  This
has been confirmed by a more detailed analysis of the dust opacity of LSBG 
disks
in comparison to those of high surface brightness galaxies, where LSBGs appear
effectively transparent (Holwerda et al.\  2005).
  
Additionally,
Pickering \& van der Hulst (1999) attempted to detect dust in LSBGs using
submillimeter observations from the JCMT with SCUBA.  Ten galaxies were 
observed, two of which were detected at 850\,$\micron$ with only one
detected at 450\,$\micron$.  They combined their submillimeter data with 
existing {\it IRAS} data, finding dust temperatures in the range 15-20\,K.
None of their very LSB ($\mu_B \le 23.5$) galaxies were detected, and
they concluded that the lack of detection in the lowest surface brightness
galaxies was consistent with previous lines of evidence that only
modest amounts of dust could exist.

{\it Spitzer} opens the opportunity to study any dust that may exist
at low levels in LSBGs.  The increased sensitivity relative to previous 
observations 
gives a higher likelihood of the detection of diffuse dust emission that 
echoes the LSBGs' diffuse optical appearance.  In addition, the resolution 
of the {\it Spitzer} imaging instruments allows for analysis of dust 
temperature, mass, and spatial distribution in LSBGs not possible before,
making it feasible to address crucial issues regarding chemical evolution
and dust production.  We present here the infrared properties
of a small sample of LSBGs with the full suite of {\it Spitzer} instruments.

\section{OBSERVATIONS}
All {\it Spitzer} observations presented here are part of the Dust
in Low Surface Brightness Galaxies Guaranteed Time Observation
Program (P.I.D. 62) whose Principal Investigator is M. J. Rieke.

\subsection{Sample}
The galaxies in this sample were selected to be some of the 
brightest and closest known LSBGs, many taken from the work
of Pickering (1998).  Two galaxies are high surface brightness, low metallicity
isolated late-type spirals that are included for the purpose of 
comparison with the LSBGs.  Table 1 summarizes general
information for each galaxy, with the LSBGs and high surface brightness
galaxies separated by a line.  The objects were 
chosen such that the infrared background cirrus was low, improving 
the chances of detection of the faintest emission associated with each
galaxy, and with sufficiently large angular diameters for resolution
with the {\it Spitzer} instruments.  Here we briefly outline their main
properties and unique characteristics.

\subsubsection{Low Surface Brightness Galaxies}

{\it Malin\,1} is one of the best known LSBGs (Bothun et al.
1987; see Barth 2007 for a more recent view) and the largest gas-rich 
galaxy found to date.  Its optical disk
is six times bigger than that of the Milky Way.  Its spectrum is 
dominated by its old, metal-rich stellar population, with a smaller
contribution from hot, young stars (Impey \& Bothun 1989).  This LSBG
is the most distant object in our sample.

{\it UGC\,5675}:  This Magellanic dwarf spiral (e.g., Schneider
et al. 1990; McGaugh et al. 1995) has a very low surface brightness disk 
(Schombert \& Bothun 1988) and has the faintest $M_B$ of our sample.
It does not have any regions of distinct star formation or 
an extended H\,{\sc i} envelope (van Zee et al. 1995).

{\it UGC\,6151} is another Magellanic spiral with a small
core of optical emission surrounded by faint diffuse emission with little
structure.  McGaugh et al. (1995) note that the galaxy contains ``quite
a few faint H\,{\sc ii} regions''.

{\it UGC\,6614} has a star-forming distinctive ring and 
has a particularly extended disk that
can be traced to at least 130$\arcsec$ (van der Hulst et al. 1993).  It
is the most metal-rich LSBG known and 
is considered to be a ``cousin'' to Malin 1 due to their similar properties.
The H\,{\sc i} data show that there is a central depression
in the gas distribution, which has led to discordant flux measurements in the
literature (Bothun et al. 1985; Giovanelli \& Haynes 1989; Pickering et al. 
1997; Matthews et al. 2001), although the most recent of these works agree.

{\it UGC\,9024}:  This galaxy has a very low surface brightness disk
coupled with a normal bulge.  It has been suggested that the large
disk plus bulge indicates that it is a transition object between average
sized LSBGs with no bulge and giant Malin 1 type galaxies (McGaugh et al.
1995).

\subsubsection{High Surface Brightness Galaxies}

{\it UGC\,6879}:  This object, while in the Impey et al. (1996) catalogue
of low surface brightness galaxies, does not qualify as an LSBG, 
due to its bright central surface brightness.  (It is possible that
it was either a candidate LSBG, later observed and found to be too
bright, or that it qualified as a type of LSB disk using a ``diffuseness''
index involving the disk scale length; see Sprayberry et al. 1995 for
details).  This galaxy is one of the few in the program to have been detected 
by {\it IRAS}, with photometry measurements at 60 and 100\,$\micron$ and 
upper limits at 12 and 25\,$\micron$.  It also has a detection at
850\,$\micron$ by SCUBA (Pickering \& van der Hulst 1999), making it an 
excellent
candidate for further study in the mid- and far-IR at the higher sensitivity 
and resolution of {\it Spitzer}.

{\it UGC\,10445} is a nearby starbursting dwarf spiral.  
The {\it Spitzer} images of this object have been
examined in detail in Hinz et al.\ (2006).  It also has available {\it IRAS} 
and {\it ISO} measurements.

\subsection{Data}

Each galaxy was observed with the Infrared Array Camera (IRAC; Fazio et al.\ 
2004) at all four wavelengths
(3.6, 4.5, 5.8, and 8.0\,$\micron$), in one 5$\arcmin \times 5\arcmin$ field 
with a frame time of
30\,s and with a dither of five positions Gaussian, for a total integration
time of 150\,s per object per wavelength.  
IRAC data were reduced at the 
{\it Spitzer} Science Center (SSC) with the S14.0.0 pipeline.  
The IRAC spatial resolution 
is $\sim$\,2$\arcsec$ in all bands.

The Multiband Imaging Photometer for {\it Spitzer} (MIPS; Rieke et al.\ 2004)
data for the sample were observed in photometry mode at 24, 70,
and 160\,$\micron$.  The integration times on all galaxies were 690\,s at
24\,$\micron$, 252\,s at 70\,$\micron$, and 42\,s at 160\,$\micron$.  The MIPS
data were reduced using the Data Analysis Tool version 
3.00 (DAT;  Gordon et al.
2005), the same techniques that are used to calibrate the instrument itself.  
Two independent reductions were carried out to test for systematic
errors on these faint sources; the results agreed closely for both reductions.
The MIPS spatial resolutions are 6$\arcsec$, 18$\arcsec$, 
and 40$\arcsec$ at 24, 70, and 160\,$\micron$, respectively.  
Dates for the IRAC and MIPS observations are given in Table 1.
The MIPS 24\,$\micron$ observations for UGC\,5675 appear to be corrupted.
The observations performed by {\it Spitzer} just prior to this object
were deep, high-redshift observations and unlikely to introduce saturated
objects that might have effected our data set.  However, the southwestern
corner of almost all the 24\,$\micron$ data collection events have highly
saturated pixels marked as NaN.  Further processing and mosaicing of the image 
was impossible, and no information could be recovered.

Infrared Spectrograph (IRS; Houck et al.\ 2004) staring-mode
observations were scheduled for only three of the targets in this
program:  Malin\,1, UGC\,6879, and UGC\,10445.  
Galaxies were observed in the low resolution 5.2-8.7\,$\micron$ (SL2) and
7.4-14.5\,$\micron$ (SL1) modes and in the low resolution 
14.0-21.3\,$\micron$ (LL2) and 19.5-38\,$\micron$ channels (LL1).  
Details of the observations, including dates, integration times and
slit position angles, are given in Table 2.
The faint, diffuse, and complex nature of these
sources makes such observations difficult.  {\it Spitzer's} onboard 
peak-up algorithm centroided on a source considered far from the intended 
target position for UGC\,10445 and centroided on a bad pixel rather 
than Malin 1.  Data were reduced with
version S13.2.0 of the SSC data pipeline, with extractions using the
pipeline developed by the Formation and Evolution of Planetary Systems (FEPS) 
Legacy team (Hines et al.\ 2006).
%5.2-8.7 ramp dur 60 # cycles 14
%7.4-14.5 ramp dur 60 # cycles 14
%14-21.3 ram dur 30 # cycles 6
%19.5-38 ramp dur 30 # cycles

Additional data are available for UGC\,6879.  These include the far
and near-ultraviolet (1350-1750\,\AA, $\lambda_{eff}=1516$\,\AA, and 
1750-2750\,\AA, $\lambda_{eff}=2267$\,\AA, respectively)
images from {\it Galaxy Evolution Explorer} ({\it GALEX}; Martin et al. 2005)
made available through the NASA Extragalactic Database (NED).

\section{ANALYSIS}

\subsection{Morphology}

Figure 1 shows an image of each galaxy in the sample at all the 
{\it Spitzer} wavelengths, with the exception of UGC\,10445, which
was presented by Hinz et al.\ (2006).

IRAC detects all galaxies in the sample at 3.6 and 4.5\,$\micron$ and 
all except for UGC\,5675 at
5.8 and 8.0\,$\micron$.  In general, the LSBGs are easily detected
at the two shorter wavelength bands, representing the old stellar
population, and difficult to image at 5.8 and 8.0\,$\micron$,
with many having only point-like detections at the longer wavelengths.

The MIPS observations of the LSBGs are varied.  
There are clear detections of all LSBGs
at 24\,$\micron$ (with the exception of the corrupted data of UGC\,5675, 
as explained above), three detections
at 70\,$\micron$, and two at 160\,$\micron$.  Of these
detections, two have extended emission at 24\,$\micron$ while two have 
point-like morphologies.  At 70\,$\micron$, two are extended, with one 
point-like, and at 160\,$\micron$ two are extended, with no point-like
detections.  A summary of this rough classification is given in Table 3,
with LSBGs listed above the solid line and HSBGs listed below the solid
line.

The difficulty of detecting emission at the longer wavelengths for the 
LSBGs is not simply a result of the decreased resolution.  Figure 2
shows the three MIPS images of UGC\,6614, with both the 24 and 
70\,$\micron$ images
convolved with a kernel that transforms them to the resolution of the
160\,$\micron$ data.  The kernel was created using a Fourier technique
on the MIPS PSFs generated by STinyTim (Gordon et al.\ in preparation).
The galaxy seems to be more extended at 70\,$\micron$ than at 24\,$\micron$,
consistent with the star forming ring becoming more prominent as shown
in Figure 1.  The signal to noise is too low at 160\,$\micron$ to confirm
this trend, but the image does show that the peak remains on the galaxy
center, as defined at 24 and 70\,$\micron$.
The changing brightness and morphology of the galaxy does not seem to
be related to the resolution differences.

\subsection{Photometry}

Aperture photometry was conducted on all {\it Spitzer} images.  
Additional image processing in the form
of background subtraction was first completed by subtracting a constant value 
from
each image.  The value of this constant was determined by masking stars
in the foreground, then taking an average
of all pixel values outside the aperture used for the galaxy photometry.
For galaxies where the foreground contamination was particularly high,
we used a large region surrounding the galaxy to determine a background
value that included a sampling of foreground stars.
Table 3 shows the MIPS flux density values and their associated errors,
along with the radii of apertures used.
Table 4 shows the corresponding IRAC photometry values.  
Galaxies that are undetected
at the various wavelengths have 3\,$\sigma$ upper limits listed in Tables 3 
and 4.  Upper limits were calculated from the images themselves, using
the mean value of the sky counts and adding three times the value of the
standard deviation of the sky counts.

The photometric uncertainties are estimated to be 10\% at 3.6 and
4.5\,$\micron$ and 15\% at 5.8 and 8.0\,$\micron$.  These values include
a 3\% absolute calibration uncertainty (Reach et al.\ 2005),
a contribution for scattered light in an extended source
(W. Reach, private communication), and an uncertainty due to the sensitivity
of the measurements to the background subtraction.  The contribution of
the scattered light is higher at 5.8 and 8.0\,$\micron$.  We do not
perform aperture corrections on the IRAC photometry, which in certain
limiting cases can be up to 25-30\% for the 5.8 and 8.0\,$\micron$ bands.
In our case, it is difficult to determine this correction for the mixture of 
point and extended sources seen in the IRAC images.  We mainly use the
8.0\,$\micron$ images to establish the presence of aromatic feature emission 
in our galaxies, so uncertainties of
this magnitude have no effect on our conclusions.
The MIPS flux calibration
uncertainties are 4\% at 24\,$\micron$, 7\% at 70\,$\micron$, and 12\% at 
160\,$\micron$ (Engelbracht et al.\ 2007; Gordon et al.\ 2007; Stansberry
et al.\ 2007).  Photometric uncertainties bring these values to total errors of
10\% at 24\,$\micron$, 20\% at 70\,$\micron$, and 20\% at 160\,$\micron$.

\subsection{Dust Modeling for UGC\,6879}

UGC\,6879, with its bright detections at all IR wavelengths,
can be analyzed in detail based on the mid-infrared and submillimeter 
photometry.
Figure 3 shows the spectral energy distribution (SED) for this galaxy,
including {\it GALEX}, 2MASS, IRAC, {\it IRAS}, {\it Spitzer}, and 
SCUBA data points.  
The emission by dust at the longer wavelengths can be modeled by an equation 
of the form

\begin{equation}
F_{dust}(\lambda) = \sum C_{i} \kappa_{i}(\lambda)
B_{\lambda}(T_{D,i})
\label{eq:dustfit}
\end{equation}

\noindent
where $C_{i}=M_{dust,i}/D^{2}$ ($D\sim$\,32\,Mpc), $\kappa_i$ is the mass
absorption coefficient, $B_{\lambda}$ is the Planck
function, $M_{dust,i}$ is the dust mass, and the sum extends
over the number of dust components.  We adopt a two-component dust
model consisting of warm and cool silicate grains ($a\sim0.1\,\micron$).
Further details regarding model assumptions and the fitting process 
can be found in 
Hinz et al. (2006).  The data set is best fitted by a model consisting of a
warm silicate component at T$=51.51^{+1.41}_{-1.28}$\,K 
and a cool silicate component at $14.94^{+0.53}_{-0.49}$\,K, 
shown in Figure 3, where the quoted error bars are 1 $\sigma$.  We estimate
the dust masses of UGC\,6879 to be $8753^{+2469}_{-2017}$\,M$_{\odot}$ 
for the warm component and 
$3.50^{+0.63}_{-0.54}\times10^7$\,M$_{\odot}$ for the cool dust, where
the quoted error bars are again 1 $\sigma$.  As
shown in Hinz et al. (2006), choosing carbonaceous grains in place of
silicate grains only modestly affects these values.

\subsection{Spectroscopy}

Figure 4 shows the full IRS spectra for UGC\,6879, Malin\,1, UGC\,10445.
We identify emission lines clearly detected 
in UGC\,6879 and UGC\,10445:  [S\,{\sc iv}], [Ne\,{\sc ii}], and
[S\,{\sc iii}] (see, e.g., Smith et al.  2004).
Additionally, we see the broad emission features usually attributed to
polycyclic aromatic hydrocarbons (PAHs).  The data show the four main
aromatic bands at 6.2, 7.7, 8.6, and 11.3\,$\micron$.  The aromatic feature at
12.7\,$\micron$ is likely to be contaminated by the [Ne {\sc ii}] 
12.8\,$\micron$ line.

There are no spectral features detected in the Malin 1 spectrum.  
Despite the
fact that {\it Spitzer} was not aligned on the coordinates given as
the central nucleus of the galaxy, the large extent of Malin 1 ensures 
that IRS took data on some portion of the disk, and the exposure time is long. 
Also, our photometry (Tables 3 and 4) shows the mid-IR excess to be very
weak.  It is likely that no aromatic features in Malin 1 are detectable 
with IRS in reasonable exposure times.

The spectra for UGC\,6879 and UGC\,10445 were fitted with the publicly
available IDL tool PAHFIT, which was developed to decompose IRS spectra 
of PAH emission sources, with a special emphasis on the careful recovery 
of ambiguous silicate absorption, and weak, blended dust emission features
(Smith et al. 2006).\footnote{Available at http://turtle.as.arizona.edu/jdsmith/pahfit.php.}  
The spectra were first prepared for 
PAHFIT by eliminating points with negative flux or with low ratios ($\le 2$)
of signal-to-noise.  Table 5 shows the fluxes or equivalent widths (EW) for the
various features as given by PAHFIT.  The 7.7\,$\micron$ complex is a 
sum of the 7.4, 7.6
and 7.9\,$\micron$ features.  PAHFIT does not calculate uncertainties on
equivalent widths because it is difficult to compute uncertainities on
the continuum of the spectrum.  The errors given in Table 5
assume that the fractional errors on the equivalent widths are the same
as on the integrated features, and thus are lower limits. 

\section{DISCUSSION}

\subsection{Comparison of UGC\,6879 and UGC\,10445}

The temperature of the cool dust, T$\sim15$\,K, found for UGC\,6879,
a high surface brightness spiral, 
coincides with that found for the starbursting dwarf galaxy UGC\,10445 
(Hinz et al.\ 2006) using similar data and modeling techniques.
It is also in agreement with the submillimeter temperature estimates
of such dust in LSBGs by Pickering \& van der Hulst (1999) and with
infrared and submillimeter estimates of the temperatures of other low
metallicity environments such as dwarf galaxies (Popescu et al.\ 2002;
Lisenfeld et al.\ 2002; Bottner et al.\ 2003).  Additionally, 
observations of normal-sized high surface brightness galaxies, including
the Milky Way (Reach et al.\ 1995; Lagache et al.\ 1998), 
show that cool dust components exist, and it is becoming apparent that
such a cool component is fairly ubiquitous across galaxy types
(see review by Tuffs \& Popescu 2005).  

The total calculated dust mass of UGC\,6879 of 
$\sim3.5\times10^7$\,M$_{\odot}$ falls
within the range found for normal high surface brightness spiral galaxies of 
$10^6-10^8$\,M$_{\odot}$ (e.g., Sodroski et al.\ 1997; Bendo et al.\ 2003) and
is a factor of ten higher than the mass values for UGC\,10445 
(Hinz et al.\ 2006).  The cool dust mass value for UGC\,10445 is considered
a lower limit due to the fact that MIPS data are insensitive to
dust colder than T=15-20\,K.  The SCUBA 850\,$\micron$ detection of
UGC\,6879 allows us to better estimate the turnover of the SED.
The better constrained fit puts somewhat tighter constraints on the dust
mass. 
%This
%may account for the factor of 10 difference found in the cool dust masses
%of UGC\,6879 and UGC\,10445.  

%The star formation rate for UGC\,6879 is lower than that for UGC\,10445,
%at 0.01\,M$_{\odot}$ yr$^{-1}$ (Burkholder et al. 2001) versus 
%0.25\,M$_{\odot}$ yr$^{-1}$ (Hinz et al. 2006).  More here...

The H\,{\sc i} gas mass to dust mass ratio found for UGC\,10445 was
$\sim$\,500 (Hinz et al. 2006).  This was found to be inconsistent with the mean value 
of the ratio for normal spiral galaxies ($71\pm49$; Stevens et al.\
2005), although the uncertainty in the dust mass value was large.
The total H\,{\sc i} mass for UGC\,6879 is $1.10\times10^9$\,$M_{\odot}$
(Sauty et al. 2003), giving a H\,{\sc i} gas mass to dust mass ratio
%of $\sim$\,30.  This value is consistent with the Stevens et al. (2005)
of $31_{-5}^{+6}$.  This value is consistent with the Stevens et al.\ (2005)
mean value.

\subsection{Comparison Between Low and High Surface Brightness Galaxies}

Popescu et al. (2002) propose that cool dust in galaxies is heated by the 
diffuse non-ionizing ultraviolet radiation produced by young stars, with a 
small contribution from the optical radiation produced by old stars.  This 
appears to be borne out for the high surface brightness galaxies, 
UGC\,6879 and UGC\,10445, where the {\it GALEX} and 
24\,$\micron$ images pinpoint the active star formation sites, and the
corresponding 160\,$\micron$ emission traces the detectable cool dust.  
Figure 5 shows the central $B$-band surface brightnesses for the sample 
versus the ratio of 24\,$\micron$ to 160\,$\micron$ flux density.  The lower 
the 
central optical surface brightness for each object, the
lower the this ratio appears to be.  This implies that there are not
large amounts of dust extinction; no
highly obscured star formation is uncovered
at 24\,$\micron$, and those galaxies with the lowest surface
brightnesses, i.e., without bright star-forming regions, are not detected at 
160\,$\micron$.  
%The outlier
%in both panels of Figure 4 is UGC\,6614; this galaxy is the most metal-rich
%LSBG known (consistent with its ``normal'' H\,{\sc i} mass to dust mass
%ratio), has a distinct star forming ring (again, unusual for
%LSBGs), and also has odd radio properties, including a central
%depression in its H\,{\sc i} emission (Pickering 1998).  

The appearance of broad aromatic emission spectral features in the isolated 
star-bursting galaxies confirms the presence of dust grains indicated 
by the IRAC, MIPS, {\it IRAS} and SCUBA images and photometry and our dust 
modeling.  Aromatic emission is believed to originate mostly from 
photodissociation envelopes at the edges of star-forming regions that
are bathed in ultraviolet photons, with some suggestion that B stars no
longer associated with H\,{\sc ii} regions can also contribute to the
heating (Spoon 2003; Calzetti et al.\ 2005).  In the cases of UGC\,6879 and
UGC\,10445, the high surface brightness galaxies for which we have IRS 
spectra, we clearly see star formation 
regions indicated by bright clumpy regions in the 24\,$\micron$ images
and the corresponding 8\,$\micron$ emission that presumably accounts for
the aromatic features.

In contrast, we see that the LSBG Malin\,1 does not have dust 
emission at far-IR wavelengths, nor aromatic emission, which is 
not surprising, given the Popescu et al.\ (2002) model and explanation. 
Malin 1 exhibits no active star-forming regions detectable at any of the
wavelengths that indicate such activity.  Without those regions,
UV photons cannot heat any existing dust to emit at long wavelengths,
nor can the envelopes believed to be the site of aromatics exist. 
That is not to say that dust cannot exist in such an object, but simply
that any such dust will not be heated and will not be detectable in far-IR
images.  This appears to be consistent with results for
irregular dwarf galaxies, where aromatic emission is found only in the 
brightest H\,{\sc ii} regions or where there is widespread, intense star
formation (Hunter et al.\ 2006).  
Braine et al.\ (2000) calculated an average star formation
rate over a lifetime of $10^{10}$\,yr
for Malin\,1 of $5 M_{\odot}$ yr$^{-1}$ based on its $V$-band
luminosity.  From this value they used Scoville \& Young (1983) to 
derive a far-IR luminosity and translated this to an expected flux density of
$\sim$\,100\,mJy at the {\it IRAS} 100\,$\micron$ band.  This was below the 
detection limit of {\it IRAS} and indeed was not detected. 
The longer integration times with MIPS now place that one-sigma
upper limit at 160\,$\micron$ of $\sim$\,10\,mJy, with Malin 1 still invisible.
One explanation for this low IR luminosity is that the current star
formation rate is far below the average over the life of the galaxy.

Additionally, simply scaling the two-component dust model for high surface
brightness galaxy UGC\,6879
down to the 24\,$\micron$ flux density values for the LSBGs does not
appear to fit their SEDs.  A scaled model that fits, for instance, a 
24\,$\micron$ flux density of 0.018\,Jy, would predict a 70\,$\micron$
flux density of $\sim$\,0.4\,Jy and a 160\,$\micron$ flux density of
$\sim$\,1.5\,Jy.  Comparing with the measurements of UGC\,6614 shows 70 and 
160\,$\micron$ outputs only $\sim$\,25\% of these predictions.  The 
160\,$\micron$ output of UGC\,6151 also appears to be somewhat below the
expected value.  Therefore, it appears that the emission
at the longer wavelengths for at least some LSBGs is fundamentally different 
from that of high surface brightness galaxies and that they are not simply
low-luminosity versions of normal galaxies.  Either LSBGs do not produce
or maintain dust in the same quantities as other galaxies, or the dust
is much colder and, therefore, undetectable in the far-IR.

Comparisons of the {\it Spitzer} data for different LSBGs may also reveal
differences in evolutionary history.  Figure 6 shows the IR SEDs of all the
galaxies in the sample.  While the two high surface brightness galaxies have
similar SEDs from 3.6 to 160\,$\micron$, the LSBGs show a variety of
steepnesses between wavelengths.  Some have steepnesses from 24 to 
160\,$\micron$ that are similar to the high surface brightness galaxies,
while others are shallower (UGC\,6614), and some appear to turn over after
70\,$\micron$ (UGC\,9024).  For instance, UGC\,6151 and UGC\,6614
have very similar far-IR flux densities, yet UGC\,6614 is much brighter
in red giant light, as represented by the 3.6\,$\micron$ flux density, 
compared with UGC\,6151.  This may imply that UGC\,6614 formed stars at
a much greater rate in the past, accumulating an old stellar population,
while UGC\,6151 may have formed stars at a more constant rate over its
lifetime.  The relatively high metallicity of UGC\,6614 supports the
hypothesis that its star formation was more vigorous in the past.  

\subsection{Metallicities and IR Properties of LSBGs}

LSBGs are generally metal-poor, consistent with the well-known
luminosity-metallicity ($L-Z$) relation for other galaxies (e.g.,
de Naray et al. 2004).  In Figure 7 we show the metallicities of
the entire HSBG plus LSBG sample versus the absolute magnitude at 
24\,$\micron$.  
Absolute magnitudes are calculated using the MIPS 24\,$\micron$ magnitude 
zero point of $7.17\pm0.0815$ calculated by Engelbracht et al. (2007).  
Average metallicities
are taken from a variety of sources in the literature (de Naray
et al. 2004; McGaugh 1994).  Others are calculated using the
Sloan Digital Sky Survey - Sky Server.\footnote{Available at http://cas.sdss.org/dr5/en/.} 
Equivalent widths of optical emission lines such as [N {\sc ii}]
and [O {\sc iii}] are available online, and we use those values in
conjunction with the rough metallicity formulations of 
Wegner et al. (2003) and Salzer et al. (2005) to obtain metallicities.
Metallicities are 
notoriously difficult to determine for LSBGs, and the variety of
sources used to obtain them for this sample may inflate errors.  
However, Figure 7 shows a weak correlation in the expected direction that
higher metallicity galaxies have brighter absolute magnitudes at 24\,$\micron$.
%between these two quantities is consistent with non-correlations
%of metallicity and central surface brightness seen previously (McGaugh 
%et al. 1994).

To probe the physical properties of galaxies that may contribute to
the lack of aromatic emission features, we calculate $R_1$, a comparison
of the contribution of 8\,$\micron$ flux with the shorter IRAC 
wavelengths defined as 
$(F_\nu(4.5\,\micron) - \alpha F_\nu(3.6\,\micron))/F_\nu(8\,\micron)$, 
and $R_2$, which is the ratio of the 8 and 24\,$\micron$ 
flux densities, for all galaxies
that are detected at those wavelengths (see Engelbracht et al. 2005
and their Figures 1 and 2).  We show $R_1$ versus $R_2$ and 
$R_2$ versus the metallicity of each galaxy in Figure 8.
The data points have large error bars associated with the photometry so
that trends are difficult to determine.
We see that the values for the LSBGs are consistent with
those found for normal galaxies by Engelbracht et al. (2005), 
occupying similar parameter space as their high surface brightness
counterparts in both plots.  Most of our sample have
relatively high 8-to-24\,$\micron$ flux ratios, so that the correlation
of increasing $R_2$ with decreasing $R_1$ is not sampled by our
galaxies.  In fact, all the galaxies in our sample that are detected at
both 8 and 24\,$\micron$ have $R_2$ larger than 0.2,
and all of the galaxies in the Engelbracht et al. (2005) with $R_2$ greater
than this value have detected aromatic features.
We see the same general metallicity trend as 
Engelbracht et al. (2005), with lower metallicity galaxies displaying
weak aromatic emission, that is, diminishing 8\,$\micron$ flux density relative
to 24\,$\micron$ flux density.  One explanation for this trend
is that harsh radiation fields in low-metallicity galaxies destroy PAH
molecules (Galliano et al. 2003, 2005; Madden et al. 2006).  
This is unlikely to be the case for LSBGs, where the radiation
fields are presumably not strong enough to destroy aromatics.  Another
explanation is that there are not enough carbon-rich asymptotic red-giant 
branch stars necessary to create large amounts of aromatic molecules in
low-metallicity galaxies.  

\section{SUMMARY}

{\it Spitzer} data on five low surface
brightness galaxies indicate that a fraction of these
objects contains modest amounts of dust, despite their
low metallicities and apparent transparency.  The LSBGs are detected
at all IRAC wavelengths, and two are detected at all of the MIPS
wavelengths.  Those LSBGs and late-type
high surface brightness counterparts
that have detectable dust appear to be the same galaxies that have
the largest amounts of star formation, while those that do not have
detectable dust are the most diffuse, least star-forming galaxies (e.g.,
Malin 1).
One explanation for this is that any dust existing in galaxies has to 
be heated to temperatures in the range 15-20\,K by ultraviolet photons escaping
from star-forming regions before being detectable at far-IR and 
submillimeter wavelengths.  The gathering evidence shows that modest 
amounts of dust can be created and maintained in a variety of environments 
and in galaxies of widely varying apparent formation histories.
We also find that LSBGs exhibit less far-IR
emission and greater variety in far-IR properties than is predicted by scaling 
related but higher surface brightness galaxy SEDs.

\acknowledgments

We thank Dean Hines and Jeroen Bouwman for
allowing us to use the FEPS data reduction pipeline.
This work is based on observations made with the Spitzer Space 
Telescope, which is operated by the Jet Propulsion Laboratory, 
California Institute of Technology under a contract with NASA. 
Support for this work was provided by NASA through an award 
issued by JPL/Caltech.
This research has made use of the NASA/IPAC Extragalactic Database (NED)
which is operated by the Jet Propulsion Laboratory, California Institute of
Technology, under contract with the National Aeronautics and Space
Administration.

\clearpage

\begin{deluxetable}{lllcccc}
\tabletypesize{\small}
\tablecaption{Galaxy Properties and Imaging Observation Dates}
\tablewidth{450pt}
\tablehead{
\colhead{Galaxy} & \colhead{Morph.} & \colhead{Distance} & \colhead{$\mu_{0,\rm B}$} & \colhead{M$_B$} & \colhead{Date} & \colhead{Date} \\
\linebreak & Type & (km s$^{-1}$) & (mag arcsec$^{-2}$) & & IRAC & MIPS }
\startdata
Malin\,1 & S & 24750 & 25.50\tablenotemark{a}& -22.50 & 2004 Jun 9 & 2005 Jan 30 \\
UGC\,5675 & Sm: & 1102 & 23.70\tablenotemark{b} & -12.95 & 2004 Apr 26 & 2004 Jun 2 \\
UGC\,6151 & Sm: & 1331 & 23.51\tablenotemark{c} & -17.21 & 2004 May 18 & 2004 Jun 4 \\
UGC\,6614 & (R)SA(r)a & 6351 & 24.30\tablenotemark{d} & -20.00 & 2003 Dec 19 & 2004 Jun 4 \\
UGC\,9024 & S & 2323 & 24.71\tablenotemark{e} & -16.58 & 2004 Jan 20 & 2004 Jul 10 \\
\hline
UGC\,6879 & SAB(r)d & 2383 & 20.40\tablenotemark{f} & -18.20 & 2004 Jun 9 & 2004 Jun 4 \\
UGC\,10445 & SBc & 963 & 21.79\tablenotemark{g} & -17.53 & 2004 Mar 8 & 2004 Mar 21\\
\enddata
\tablenotetext{a}{Bothun et al.\ (1987)}
\tablenotetext{b}{McGaugh \& Bothun (1994)}
\tablenotetext{c}{Patterson \& Thuan (1996)}
\tablenotetext{d}{van der Hulst et al.\ (1993)}
\tablenotetext{e}{McGaugh et al.\ (1995)}
\tablenotetext{f}{Impey et al.\ (1996)}
\tablenotetext{g}{van Zee (2000)}
\end{deluxetable}

\begin{deluxetable}{lcccccc}
\tabletypesize{\small}
\tablecaption{IRS Observation Details}
\tablewidth{350pt}
\tablehead{
\colhead{Galaxy} & \colhead{Date} & IRS Mode & \colhead{Integration Time} & \colhead{P.A.} \\
\linebreak & & & (s) & (deg) }
\startdata
Malin\,1 & 2005 Jan 4 & SL & 1707 & -160.92 \\
 & & LL & 377.5 & 115.55 \\
UGC\,6879 & 2004 Jun 27 & SL & 1707 & 19.01 \\
 & & LL & 377.5 & -64.50 \\
UGC\,10445 & 2004 Jul 14 & SL & 1707 & 42.37 \\
 & & LL & 377.5 & -41.15 \\
\enddata
\tablecomments{Complete details of observations can retrieved via the SSC's
Leopard database software.}
\end{deluxetable}

\begin{deluxetable}{lcccccccccc}
\tabletypesize{\tiny}
\tablecaption{MIPS LSBG Morphologies and Flux Densities}
\tablewidth{460pt}
\tablehead{
\colhead{Galaxy} & \colhead{Morph.} & \colhead{Morph.} & \colhead{Morph.} & \colhead{F$_{\nu}$ (Jy)} & \colhead{Radius} & \colhead{F$_{\nu}$ (Jy)} & \colhead{Radius} & \colhead{F$_{\nu}$ (Jy)} & \colhead{Radius} \\
\linebreak & \colhead{24\,$\micron$} & \colhead{70\,$\micron$} & \colhead{160\,$\micron$} & \colhead{24\,$\micron$} & \colhead{($\arcsec$)} & \colhead{70\,$\micron$} & \colhead{($\arcsec$)} & \colhead{160\,$\micron$} & \colhead{($\arcsec$)}}
\startdata
Malin 1 & point-like & no detection & no detection & $4.3$E-4$\pm$4.3E-5 & 24.90 & $< 0.004$ & \nodata & $< 0.01$ & \nodata\\
UGC\,5675 & \nodata & no detection & no detection &\nodata & \nodata & $< 0.009$ & \nodata & $< 0.02$ & \nodata\\
UGC\,6151 & extended & extended & extended & $0.005\pm5.0$E-4 & 62.25 & $0.08\pm0.02$ & 49.25 & $0.29\pm0.06$ & 80 \\
UGC\,6614 & extended & extended & extended & $0.018\pm2.0$E-3 & 62.25 & $0.08\pm0.02$ & 54.18 & $0.38\pm0.08$ & 56 \\
UGC\,9024 & point-like & point-like & no detection & $0.001\pm1.0$E-4 & 24.90 & $0.04\pm0.01$ & 24.63 & $< 0.02$ & \nodata\\
\hline
UGC\,6879 & extended & extended & extended & $0.027\pm3.0$E-3 & 62.25 & $0.44\pm0.09$ & 54.18 & $1.47\pm0.29$ & 56 \\
UGC\,10445 & extended & extended & extended & $0.025\pm2.0$E-3 & 105.8 & $0.55\pm0.11$ & 98.50 & $2.50\pm0.50$ & 120 \\
\enddata
\end{deluxetable}

\begin{deluxetable}{lcccccccccc}
\tabletypesize{\small}
\tablecaption{IRAC Flux Densities}
\tablewidth{525pt}
\tablehead{
\colhead{Galaxy} & \colhead{F$_{\nu}$ (Jy)} & \colhead{F$_{\nu}$ (Jy)} & \colhead{Radius} & \colhead{F$_{\nu}$ (Jy)} & \colhead{F${\nu}$ (Jy)} & \colhead{Radius}\\
\linebreak & \colhead{3.6\,$\micron$} & \colhead{4.5\,$\micron$} & \colhead{($\arcsec$)} & \colhead{5.8\,$\micron$} & \colhead{8.0\,$\micron$} & \colhead{($\arcsec$) }}
\startdata
Malin 1 & 1.74E-3$\pm$1.74E-4 & 1.20E-3$\pm$1.20E-4 & 18 & 6.87E-4$\pm$1.03E-4 & 1.03E-3$\pm$1.55E-4 & 18 \\
UGC\,5675 & 1.22E-3$\pm$1.22E-4 & 7.47E-4$\pm$7.47E-5 & 30 & $< 2.03$E-5 & $< 1.18$E-4 & \nodata \\
UGC\,6151 & 4.60E-3$\pm$4.60E-4 & 2.87E-3$\pm$2.87E-4 & 60 & 2.47E-3$\pm$3.71E-4 & 4.15E-3$\pm$6.23E-4 & 60 \\
UGC\,6614 & 2.45E-2$\pm$2.45E-3 & 1.43E-2$\pm$1.43E-3 & 108 & 1.66E-2$\pm$2.49E-3 & 2.43E-2$\pm$3.65E-3 & 60 \\
UGC\,9024 & 3.10E-3$\pm$3.10E-4 & 2.00E-3$\pm$2.00E-4 & 60 & 9.00E-4$\pm$1.35E-4 & 2.00E-3$\pm$3.00E-4 & 24 \\
\hline
UGC\,6879 & 2.15E-2$\pm$2.15E-3 & 1.39E-2$\pm$1.39E-3 & 108 & 2.05E-2$\pm$3.08E-3 & 4.71E-2$\pm$7.07E-3 & 60 \\
UGC\,10445 & 2.00E-2$\pm$2.00E-3 & 1.60E-2$\pm$1.60E-3 & 78 & 2.10E-2$\pm$3.15E-3 & 3.40E-2$\pm$5.10E-3 & 78 \\
\enddata
\end{deluxetable}

\begin{deluxetable}{lcccccccccc}
\tabletypesize{\small}
\tablecaption{IRS Fluxes or EWs from PAHFIT}
\tablewidth{220pt}
\tablehead{
\colhead{Feature} & \colhead{Flux}\\
\linebreak & \colhead{(erg s$^{-1}$ cm$^{-2}$)}}
\startdata
UGC\,6879 & & \\
\hline
[S \sc{iv}] & 4.89E-16$\pm$3.86E-16 \\
$[$Ne \sc{ii}$]$ & 4.98E-15$\pm$4.74E-16 \\
$[$S \sc{iii}$]$ & 7.14E-15$\pm$1.45E-15 \\
\hline
7.7\,$\micron$ complex EW & 13.57$\pm$0.68\,$\micron$ \\
\hline
& & \\
UGC\,10445 & & \\
\hline
[S \sc{iv}] & 2.36E-15$\pm$3.75E-16 \\
$[$Ne \sc{ii}$]$ & 4.20E-15$\pm$4.30E-16 \\
$[$S {\sc iii}$]$ & 2.39E-15$\pm$1.63E-15 \\
\hline
7.7\,$\micron$ complex EW & 5.69$\pm$1.83\,$\micron$ \\
\enddata
\end{deluxetable}

\clearpage

\begin{figure}
\plotone{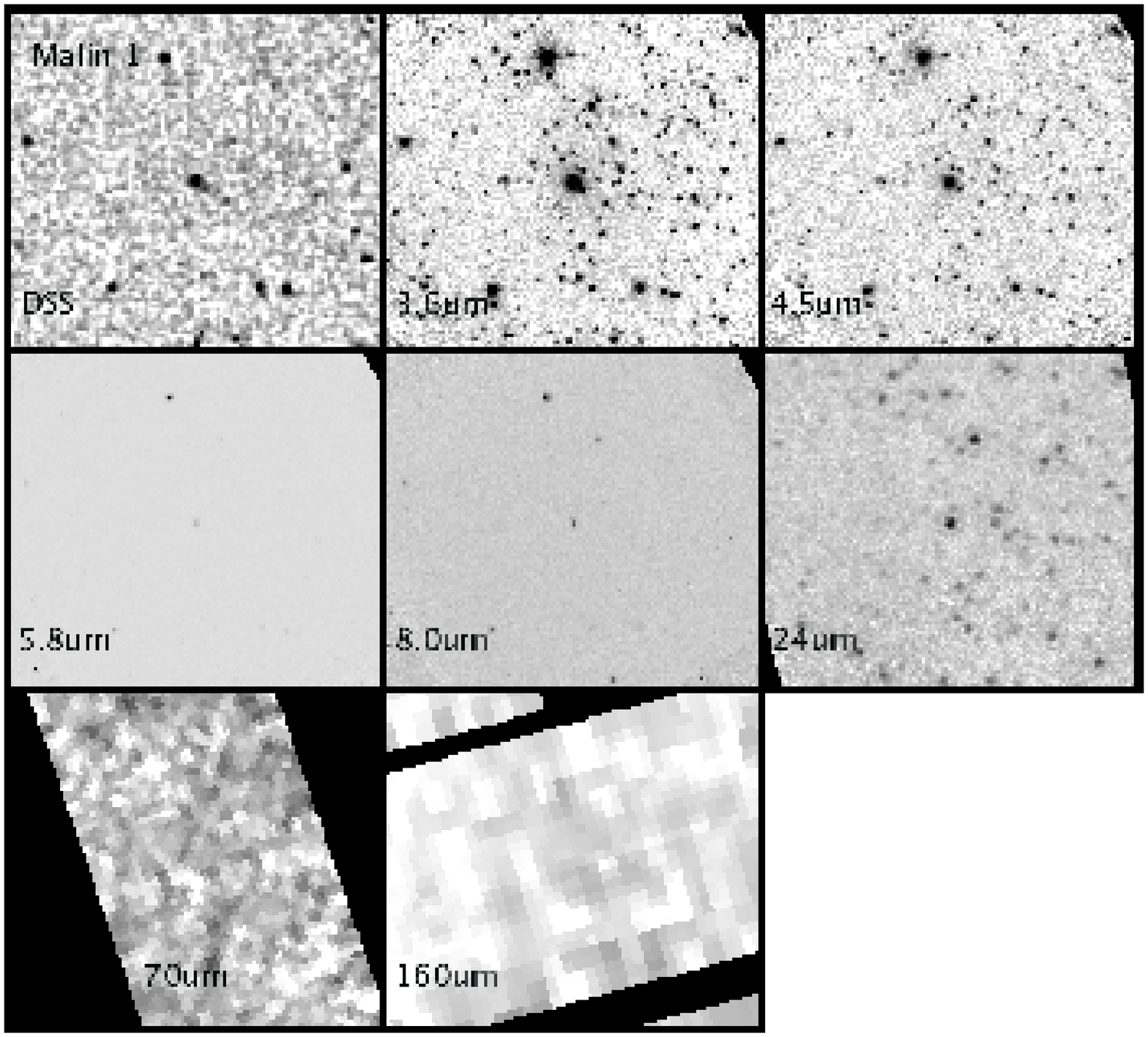}
\end{figure}

\begin{figure}
\plotone{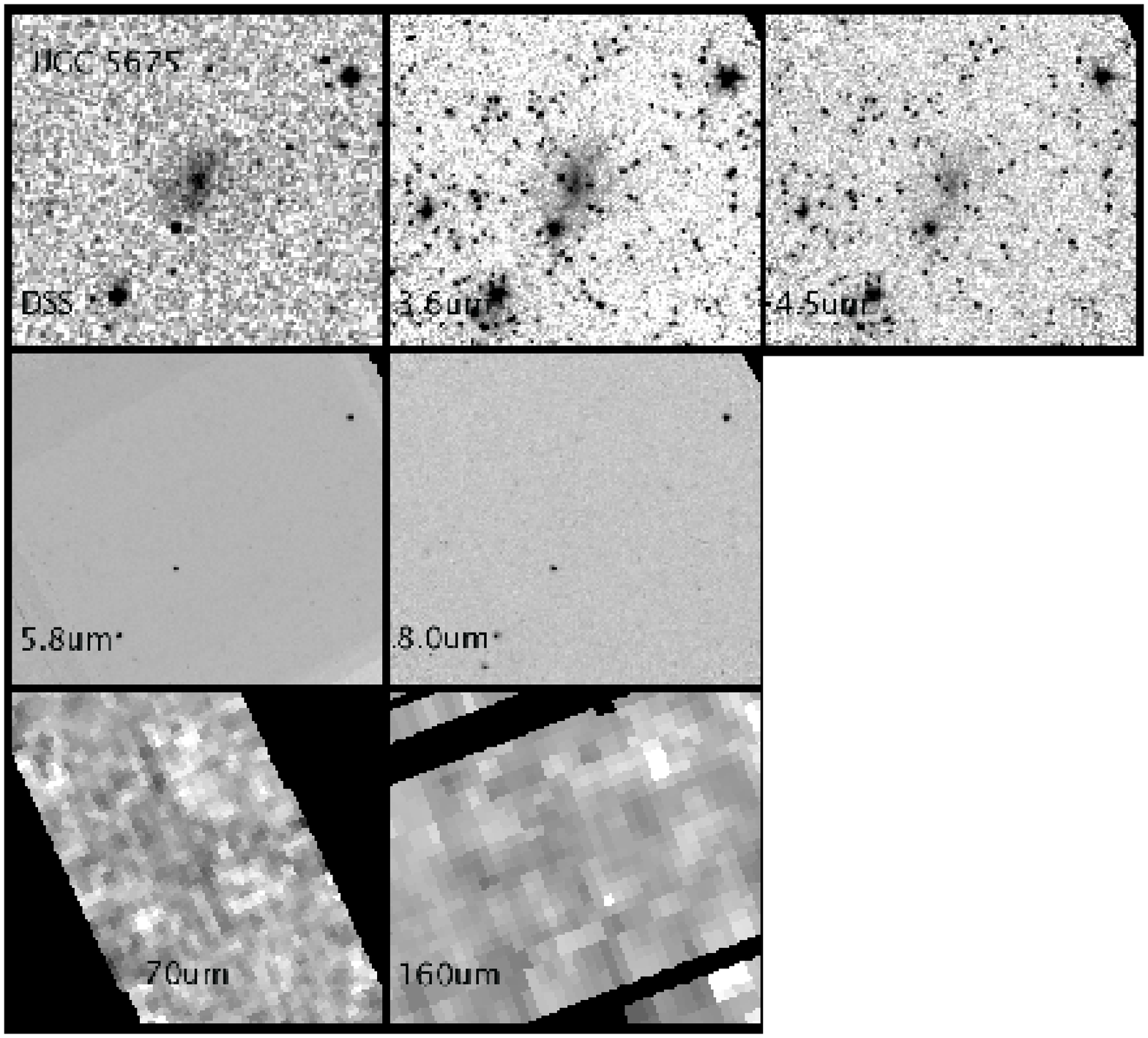}
\end{figure}

\begin{figure}
\plotone{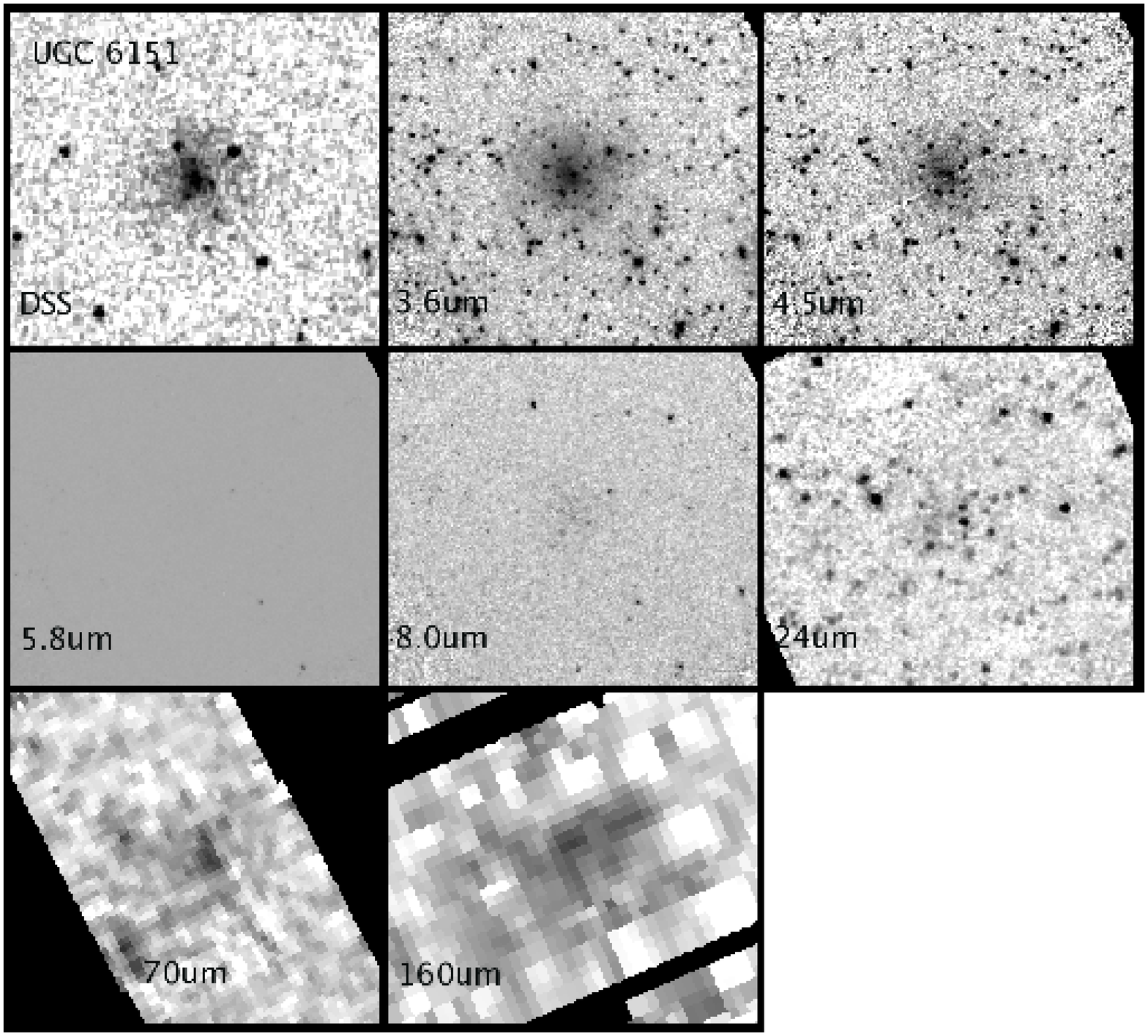}
\end{figure}

\begin{figure}
\plotone{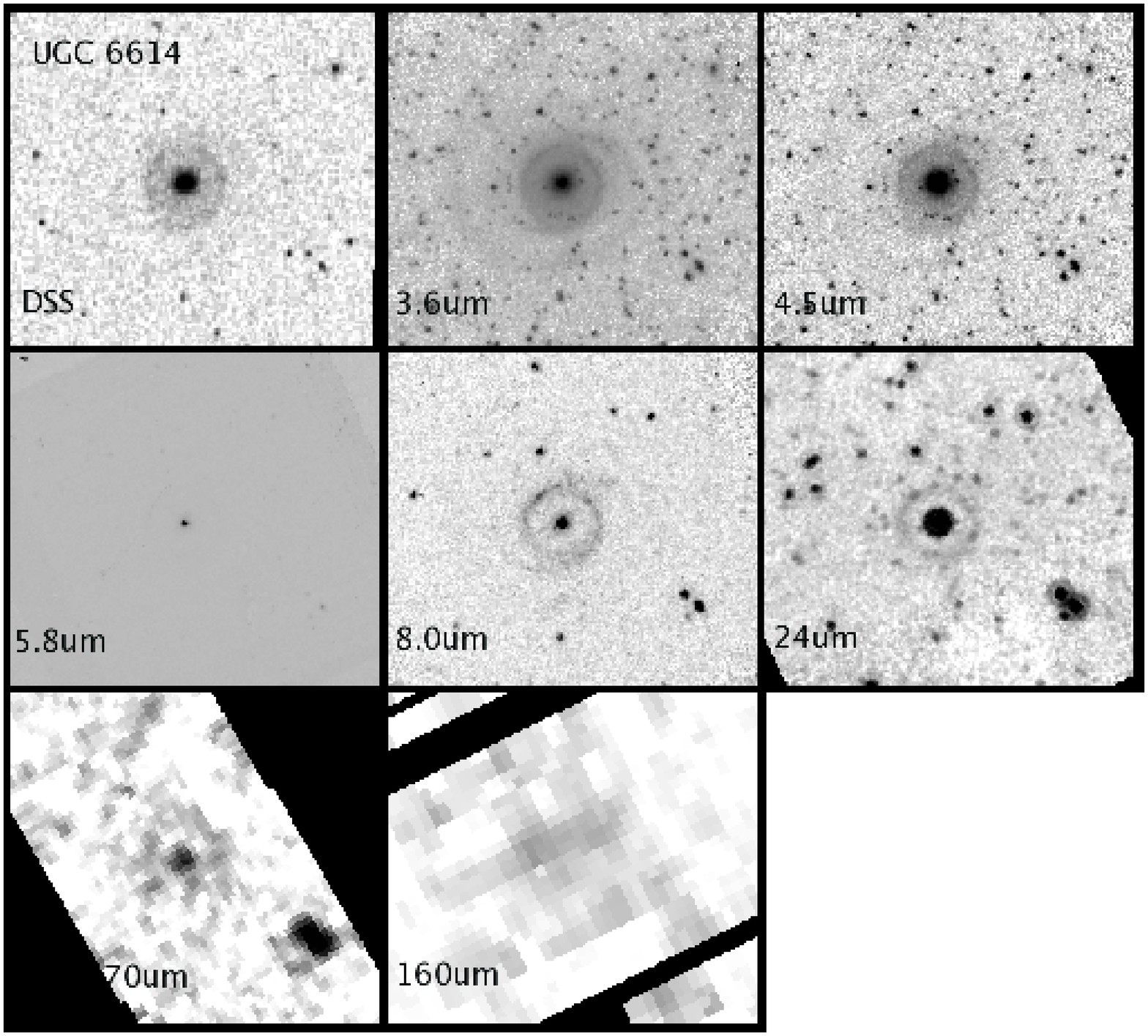}
\end{figure}

\begin{figure}
\plotone{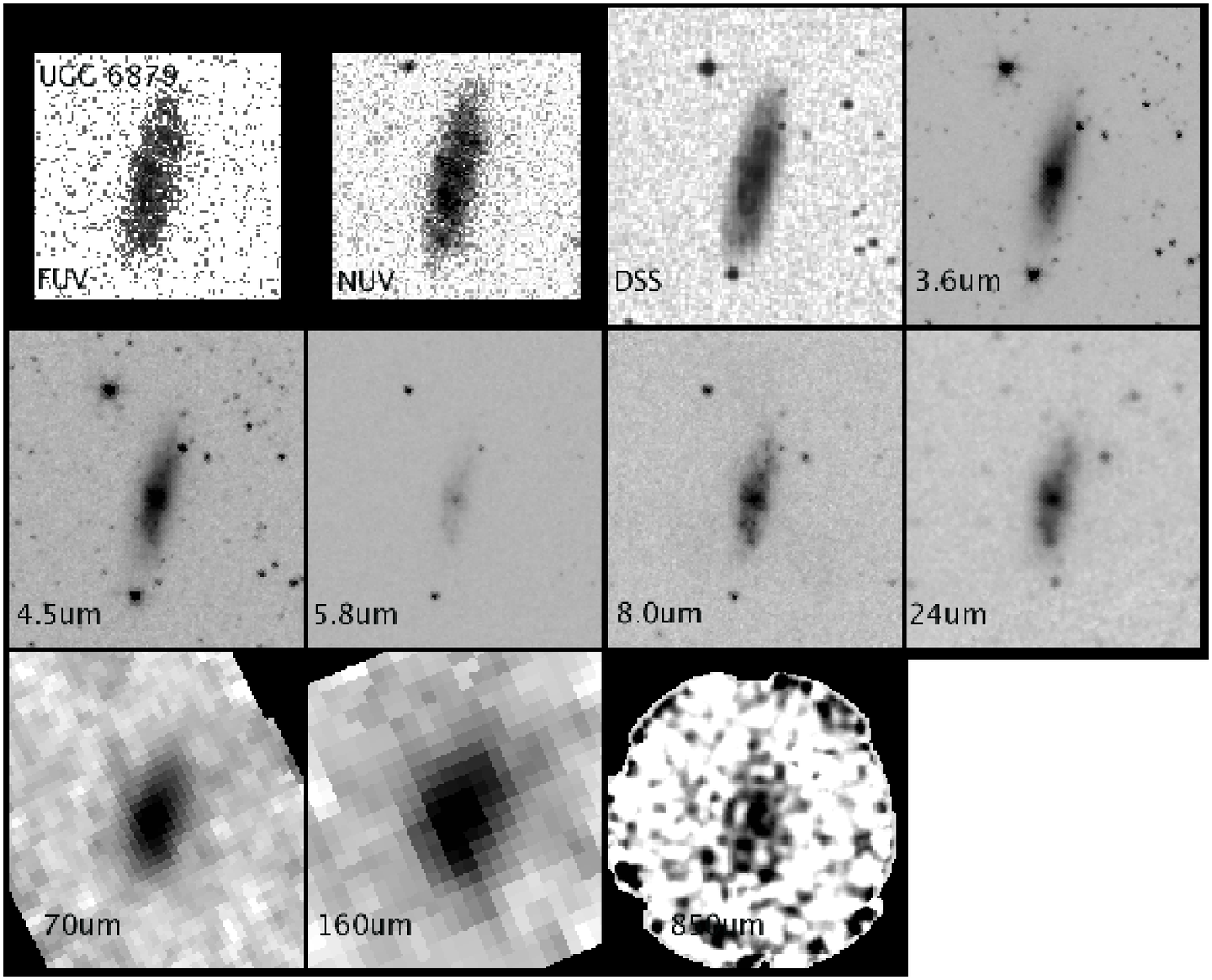}
\end{figure}

\begin{figure}
\plotone{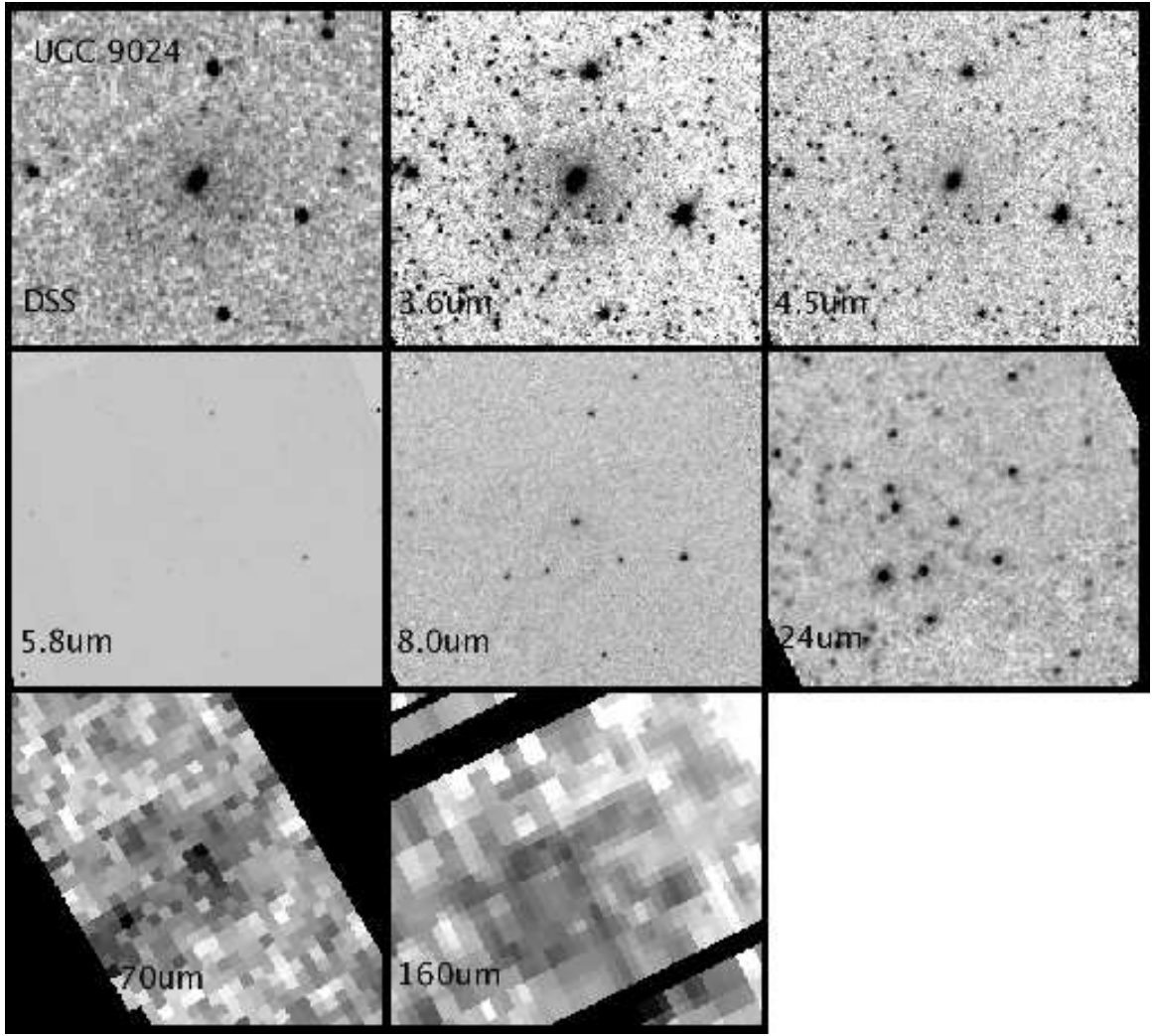}
\caption{Multi-wavelength views of the galaxy sample.  North is up and
east is to the left.  The field of view of each panel is 
$\sim4\farcm5\times4\farcm5$.  The panels, from right to left, starting
on the top row:  Digitized Sky Survey (DSS), the four IRAC bands, the three MIPS bands.
The exception to this is UGC\,6879 which has:  {\it GALEX} far-UV, {\it GALEX} 
near-UV, DSS, IRAC, MIPS, and SCUBA 850\,$\micron$.  The scale for the DSS
image is 1$\farcs$7 and for {\it GALEX} it is 5$\arcsec$.
Pixel scales for all IRAC images are 
1$\farcs$2.  Pixels scales for the MIPS images are 1$\farcs$245 for 
24\,$\micron$, 4$\farcs$925 for 70\,$\micron$, and 8$\farcs$0 for 
160\,$\micron$.  The SCUBA image has been rebinned to 1$\arcsec$ pixels
from an original beamwidth of 15$\arcsec$.}
\end{figure}

\begin{figure}
\plotone{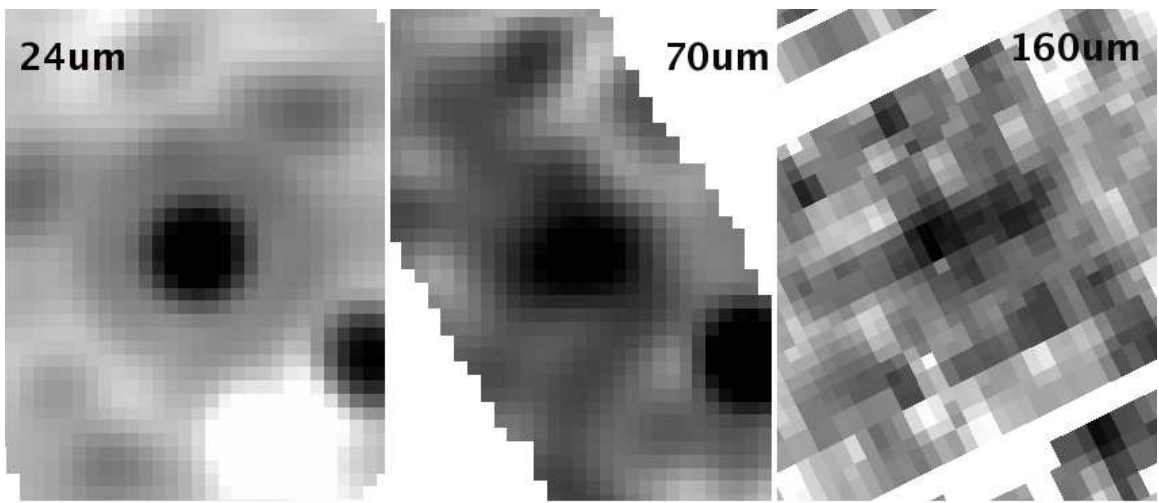}
\caption{UGC\,6614 at 24, 70, and 160\,$\micron$.  The 24 and 70\,$\micron$
images are convolved to the 160\,$\micron$ resolution.  The changing 
morphology and brightness of the galaxy at far-IR wavelengths does not
appear to be a result of the resolution differences. North is up and east
is to the left.  The field of view of each panel is 
$\sim4\farcm0\times4\farcm8$.}
\end{figure}

\begin{figure}
\plotone{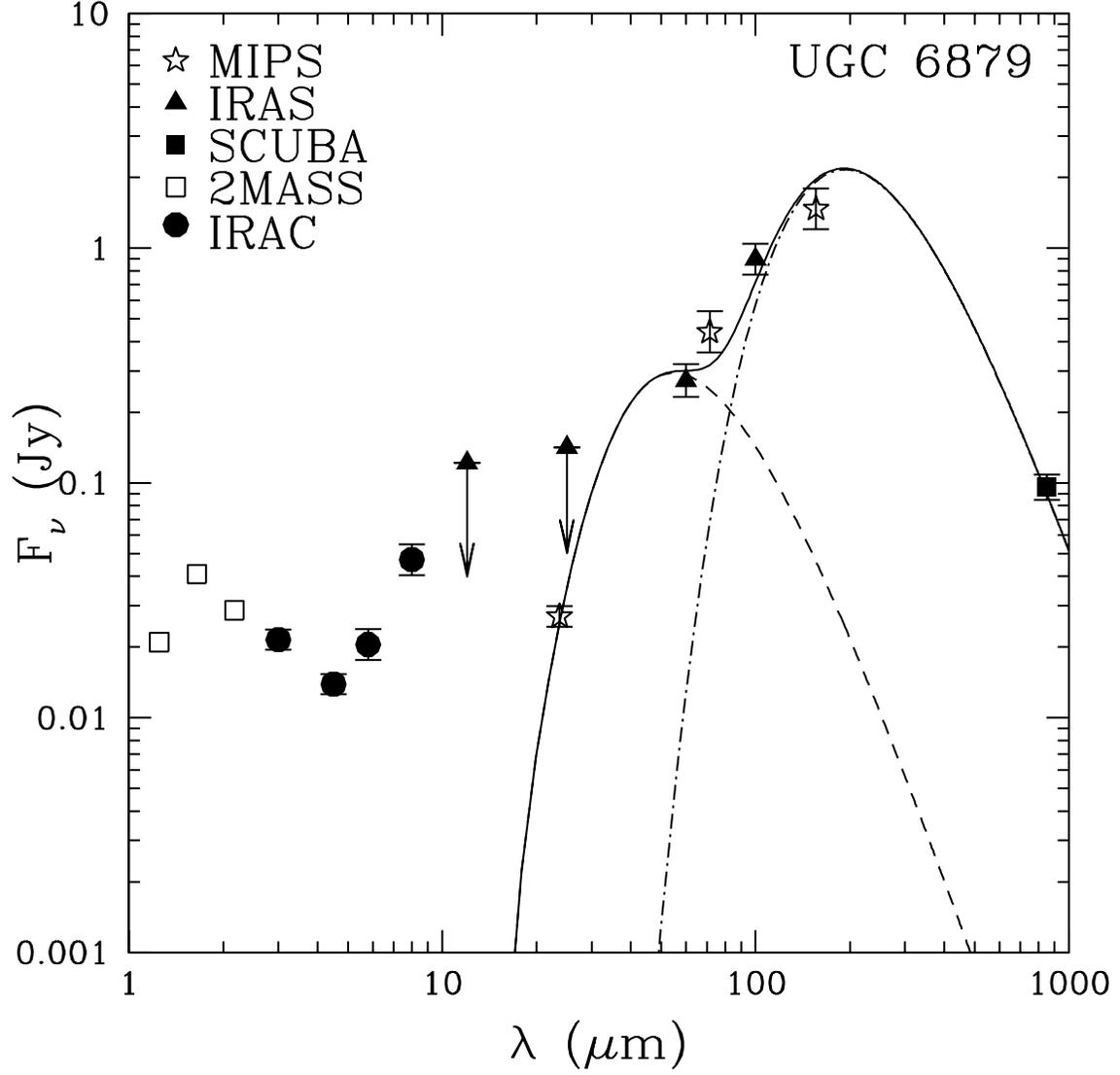}
\caption{Spectral energy distribution of UGC\,6879 showing {\it GALEX},
2MASS, IRAC, {\it IRAS}, MIPS, and SCUBA data points.  The {\it IRAS} 
data points at
12 and 25\,$\micron$ are upper limits only.  The solid line is a 
two-component dust model fitted to the four IRAC data points, two {\it IRAS}
points, and three MIPS points.  This model consists of a warm silicate
component at $T=52$\,K ({\it dashed line}) and a cool silicate 
component at $T=15$\,K ({\it dashed-dotted line}).}
\end{figure}

\begin{figure}
\plotone{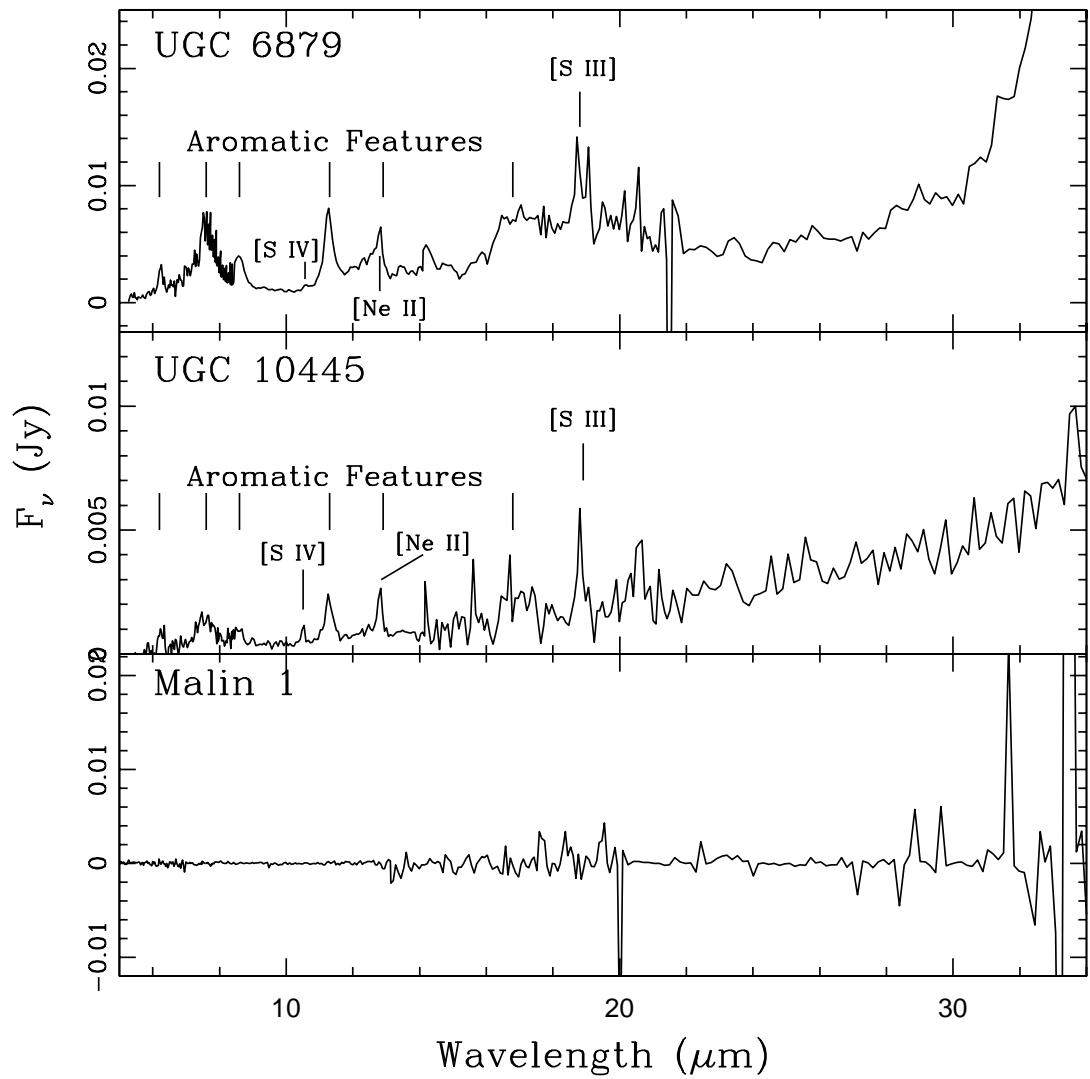}
\caption{IRS spectra, rest wavelength versus flux density, 
for UGC\,6879, UGC\,10445, and Malin 1.  
The broad aromatic features and emission lines are indicated on the top
and middle panels panel on both the UGC\,6879 and UGC\,10445 spectra.  
The Malin 1 spectrum is consistent with noise.}
\end{figure}

\begin{figure}
\plotone{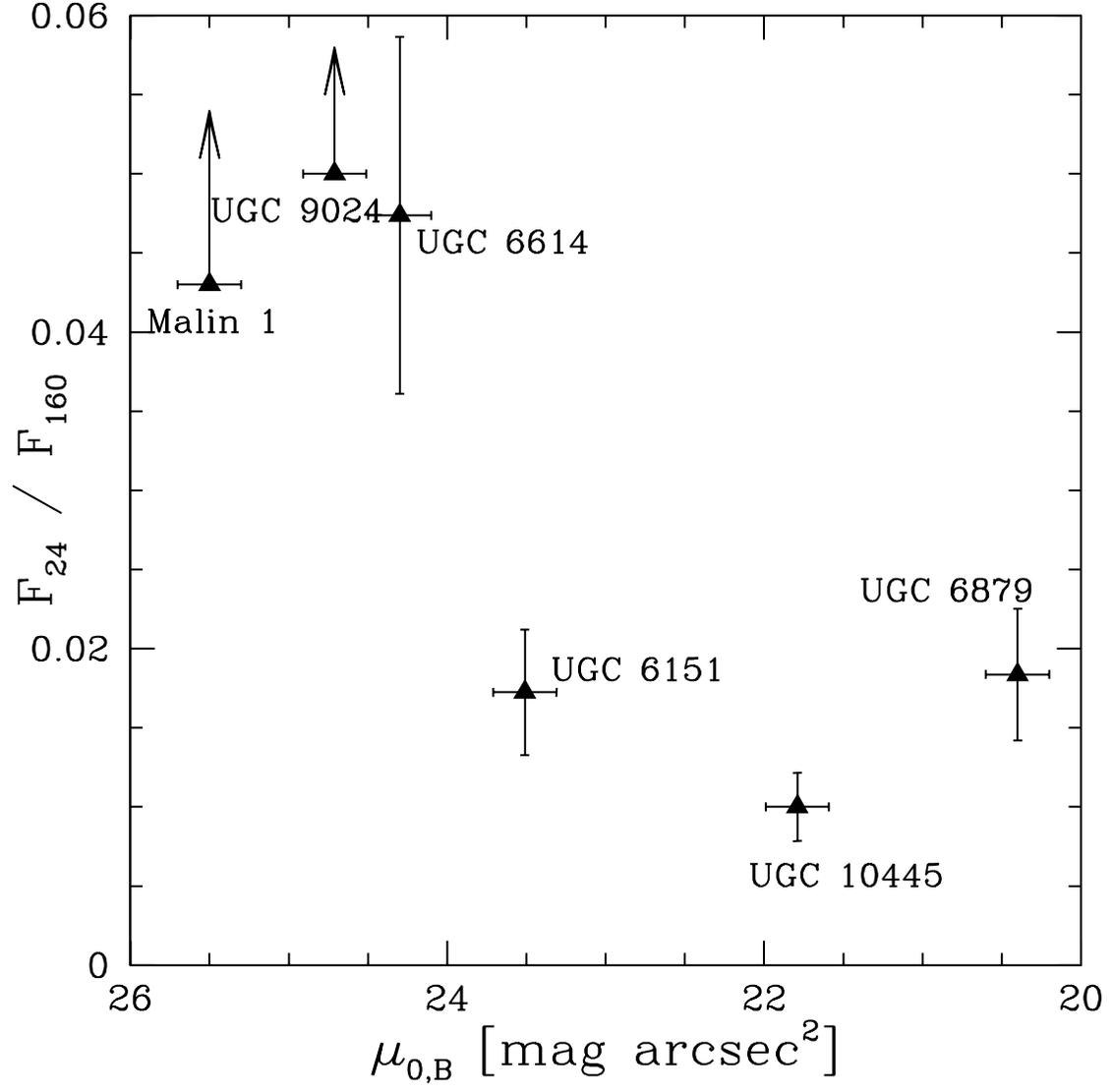}
\caption{$B$-band central surface brightnesses of the galaxy sample
versus the ratio of the flux densities at 24 and 160\,$\micron$.  All
objects are included except UGC\,5675, which does not have data at 
24\,$\micron$.  Lower limits are given for Malin\,1 and UGC\,9024,
which are not detected at 160\,$\micron$.}
\end{figure}

\begin{figure}
\plotone{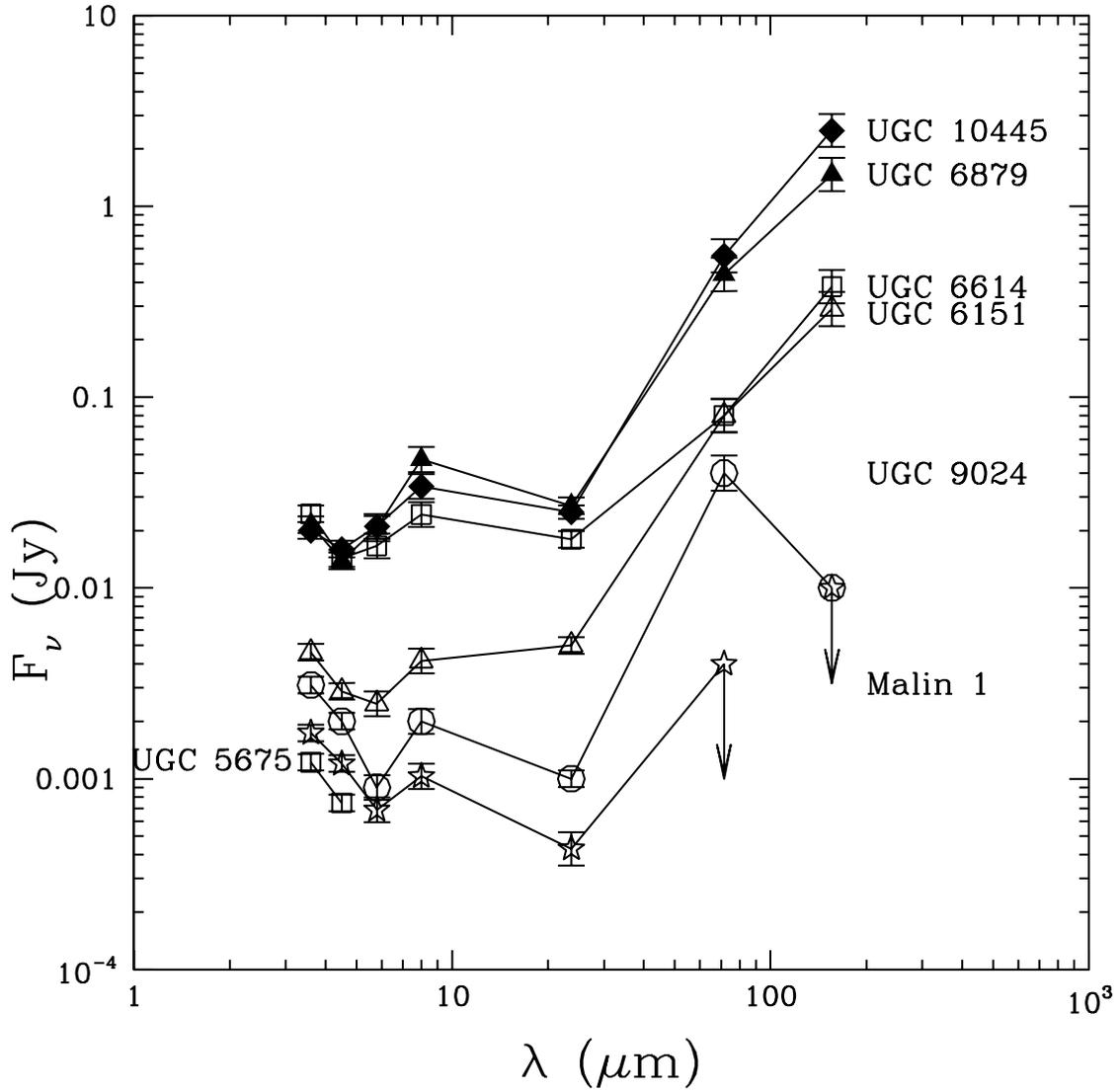}
\caption{Spectral energy distributions of all the galaxies showing
the IRAC and MIPS data points.  The high surface brightness galaxy data
are shown in solid points while the LSBG data are shown in open points.
The arrows represent 3\,$\sigma$ upper limits at 70 and 160\,$\micron$.}
\end{figure}

\begin{figure}
\plotone{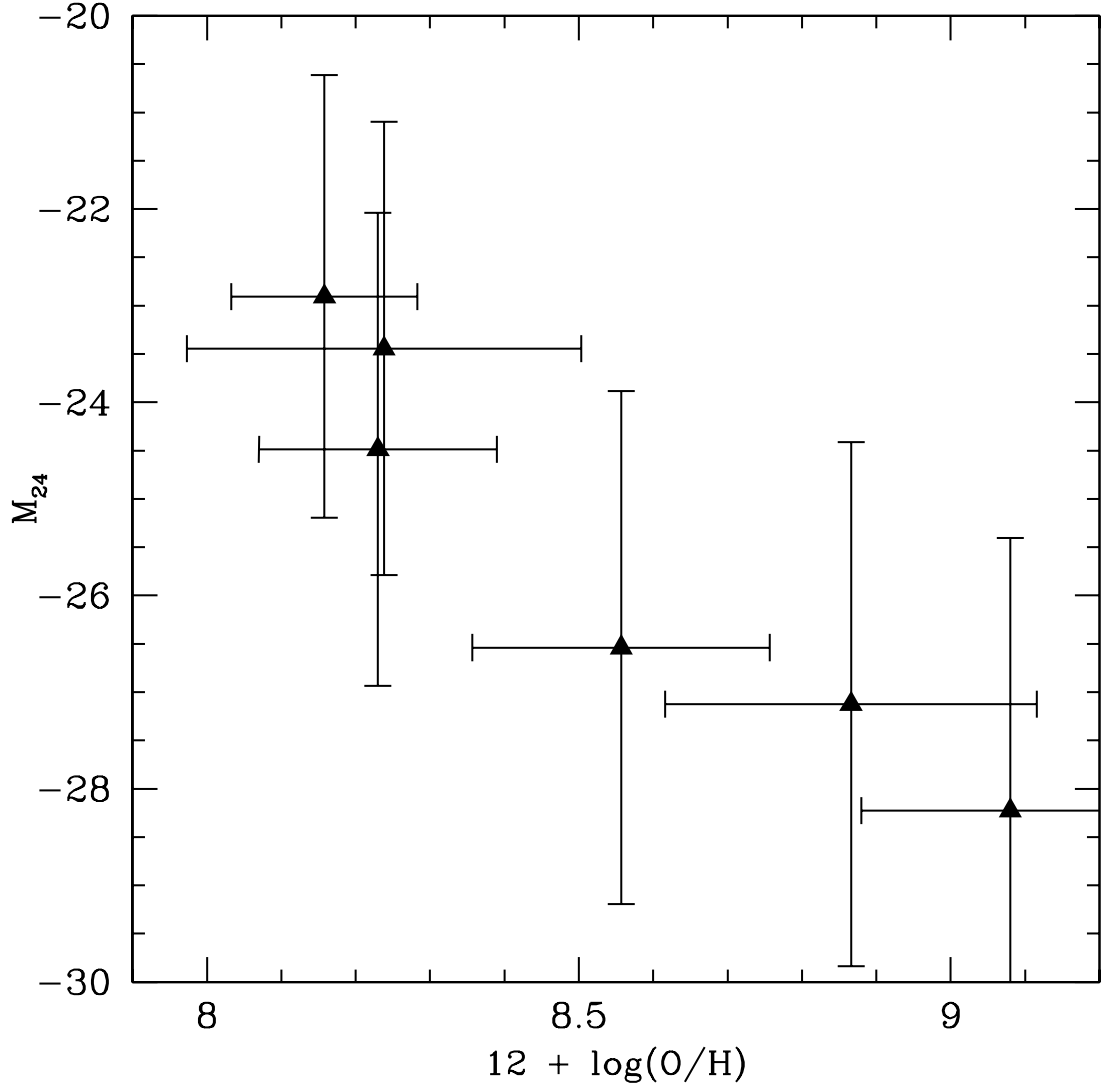}
\caption{Metallicity and absolute magnitude at 24\,$\micron$ for
all the galaxies except UGC\,5675, where 24\,$\micron$ data are
not available.  While it would
be expected for the galaxies to follow the $L-Z$ trend, the difficulty
in determining metallicities seems to have weakened the correlation.}
\end{figure}

\begin{figure}
\plottwo{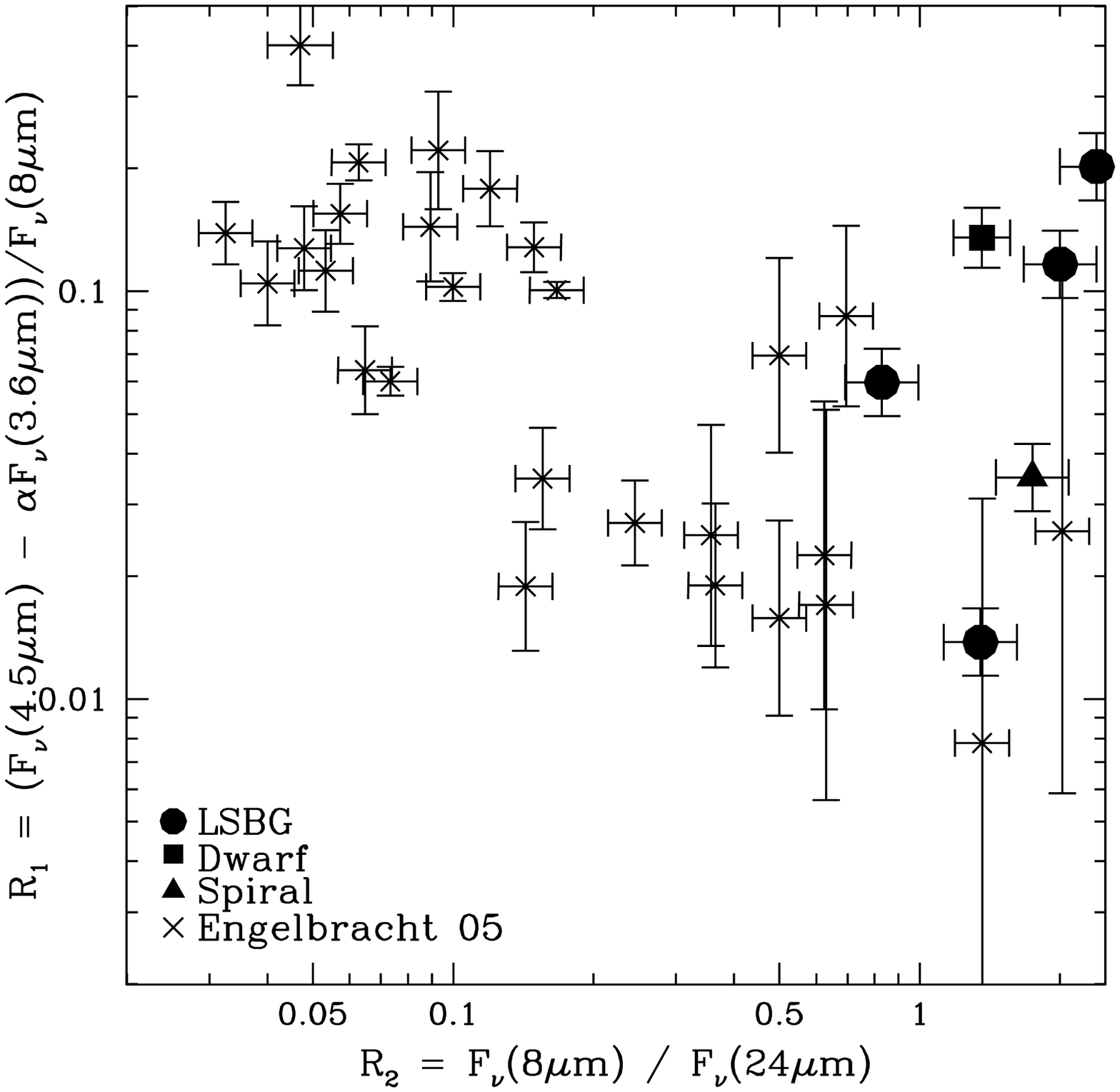}{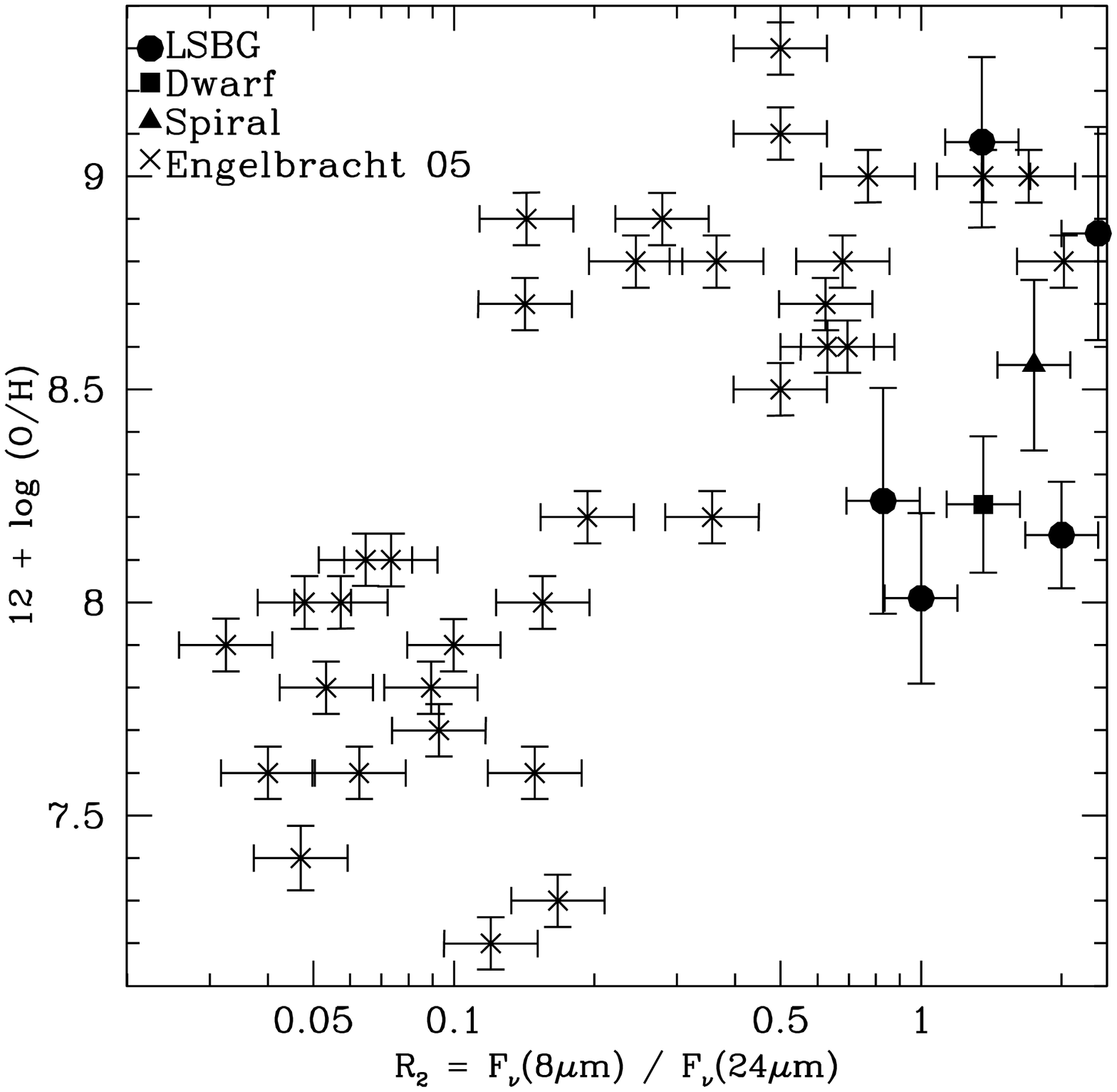}
\caption{On the left, mid-infrared colors of the galaxy sample, with $R_1$ and
$R_2$ defined as in Engelbracht et al. (2005), and, on the right, 
galaxy metallicity as a function of the 8-to-24\,$\micron$
color, $R_2$.  Solid circles represent
LSBGs, the solid square represents UGC\,10445, and the solid triangle
represents UGC\,6879.  The data points from Engelbracht et al. (2005) are
shown as crosses.  There is a slight upward trend of increasing aromatic
strength with increasing metallicity.}
\end{figure}

\end{document}